\documentclass[conference]{IEEEtran}
\IEEEoverridecommandlockouts

\usepackage{cite}
\usepackage{amsmath,amssymb,amsfonts,subfigure,bm}
\usepackage{cite}
\usepackage[numbers,sort&compress]{natbib}
\usepackage{algorithmic}
\usepackage{graphicx}
\usepackage{textcomp}
\usepackage{xcolor}
\usepackage{booktabs}
\usepackage{multirow}
\usepackage{footnote}
\usepackage{array}
\usepackage[citecolor=green]{hyperref}
\def\BibTeX{{\rm B\kern-.05em{\sc i\kern-.025em b}\kern-.08em
    T\kern-.1667em\lower.7ex\hbox{E}\kern-.125emX}}
\begin{document}

\title{MTDA-HSED:  Mutual-Assistance Tuning and Dual-Branch Aggregating for Heterogeneous Sound Event Detection\\
\thanks{ \dag \quad These authors contributed equally to this work.}
\thanks{*\quad Corresponding author.}
\thanks{\href{https://github.com/Visitor-W/MTDA}{https://github.com/Visitor-W/MTDA}} 
}

\author{
    Zehao Wang$^{1,\dag}$,
    Haobo Yue$^{1,\dag}$,
    Zhicheng Zhang$^{2,*}$, 
    Da Mu$^{1}$,
    Jin Tang$^{2}$, 
    Jianqin Yin$^{2}$ \\
    \\
    $^{1}$School of Artificial Intelligence, Beijing University of Posts and Telecommunications, China\\
    $^{2}$School of Intelligent Engineering and Automation, Beijing University of Posts and Telecommunications, China\\
    Email:\{wzhao, hby, zczhang, da.mu, tangjin, jqyin\}@bupt.edu.cn
}

\maketitle

\begin{abstract}
Sound Event Detection (SED) plays a vital role in comprehending and perceiving acoustic scenes. Previous methods have demonstrated impressive capabilities. However, they are deficient in learning features of complex scenes from heterogeneous dataset. In this paper, we introduce a novel dual-branch architecture named Mutual-Assistance Tuning and Dual-Branch Aggregating for Heterogeneous Sound Event Detection (MTDA-HSED). The MTDA-HSED architecture employs the Mutual-Assistance Audio Adapter (M3A) to effectively tackle the multi-scenario problem and uses the Dual-Branch Mid-Fusion (DBMF) module to tackle the multi-granularity problem. Specifically, M3A is integrated into the BEATs block as an adapter to improve the BEATs' performance by fine-tuning it on the multi-scenario dataset. The DBMF module connects BEATs and CNN branches, which facilitates the deep fusion of information from the BEATs and the CNN branches. Experimental results show that the proposed methods exceed the baseline of mpAUC by \textbf{$5\%$} on the DESED and MAESTRO Real datasets.

\end{abstract}


\section{Introduction}
The Sound Event Detection (SED)~\cite{Mesaros_2021} system is designed to detect sound events and corresponding timestamps (onset and offset). SED plays a vital role in comprehending acoustic scenes and perceiving physical environments, providing essential support to a wide range of related fields, such as security \cite{9187249, Nijhawan2022GunIF}, smart homes \cite{10180246, 9980350} and smart cities \cite{BonetSol2022AnalysisAA}. 

In fact, SED in the real world often have to face the problem of multi-scenario and multi-granularity. Therefore, it is crucial to explore how SED systems can leverage training data with varying annotation granularities where heterogeneous training dataset and potentially missing labels are usually considered, which is also concerned by task 4 of DCASE 2024. The official baseline\cite{cornell2024dcase} combines the Bidirectional Encoder representation from Audio Transformers (BEATs) \cite{chen2022beatsaudiopretrainingacoustic}, pre-trained on the large-scale AudioSet \cite{7952261}, with Convolutional Recurrent Neural Network (CRNN) \cite{Dinkel_2021}, leveraging the pre-trained model's prior knowledge to learn features from heterogeneous dataset. Recently, to more effectively leverage the prior knowledge of the pre-trained models for downstream tasks, fine-tuning models pre-trained on AudioSet has become a long-standing popular paradigm, such as AST-SED \cite{li2023astsedeffectivesoundevent}, PASST-SED \cite{Li2023FinetuningAS}, and ATST-SED \cite{shao2023finetunepretrainedatstmodel}. These methods represent the selective types of Parameter-Efficient Fine-Tuning (PEFT)~\cite{han2024parameterefficientfinetuninglargemodels} approaches. In addition, PEFT also includes additive methods such as ADAPTER \cite{chen2022adaptformeradaptingvisiontransformers} and MONA \cite{yin2023adapterneedtuningvisual}, which have demonstrated success in other domains. Compared to selective methods, additive methods alter the structure of the pre-trained model, enabling it to better tackle downstream tasks.

\begin{figure*}[t]
    \subfigure[BEATs Block]{
        \includegraphics[scale=0.37]{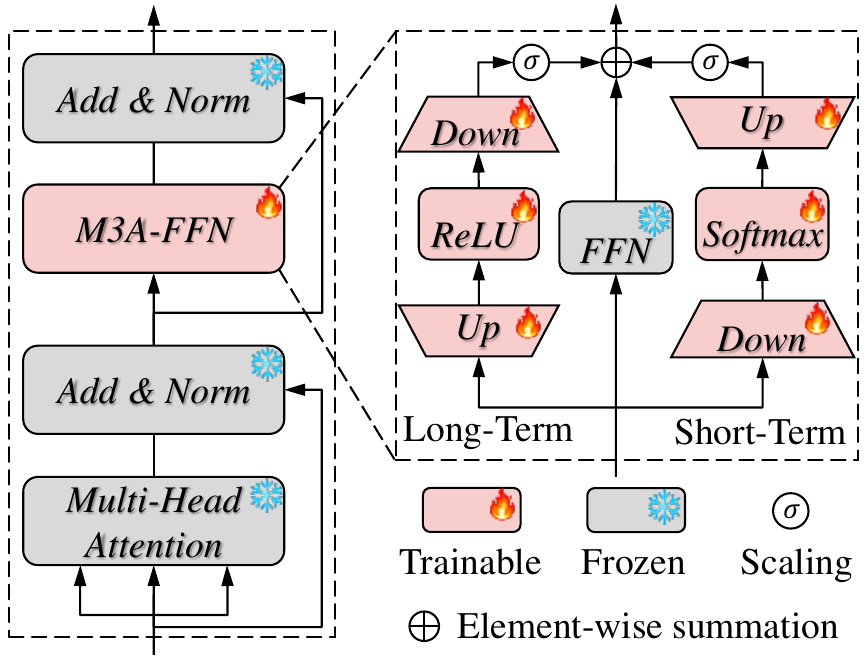}%
        \label{fig:M3A}
    }
    \hspace{-3pt}
    \subfigure[The overall architecture of MTDA-HSED]{
        \includegraphics[scale=0.37]{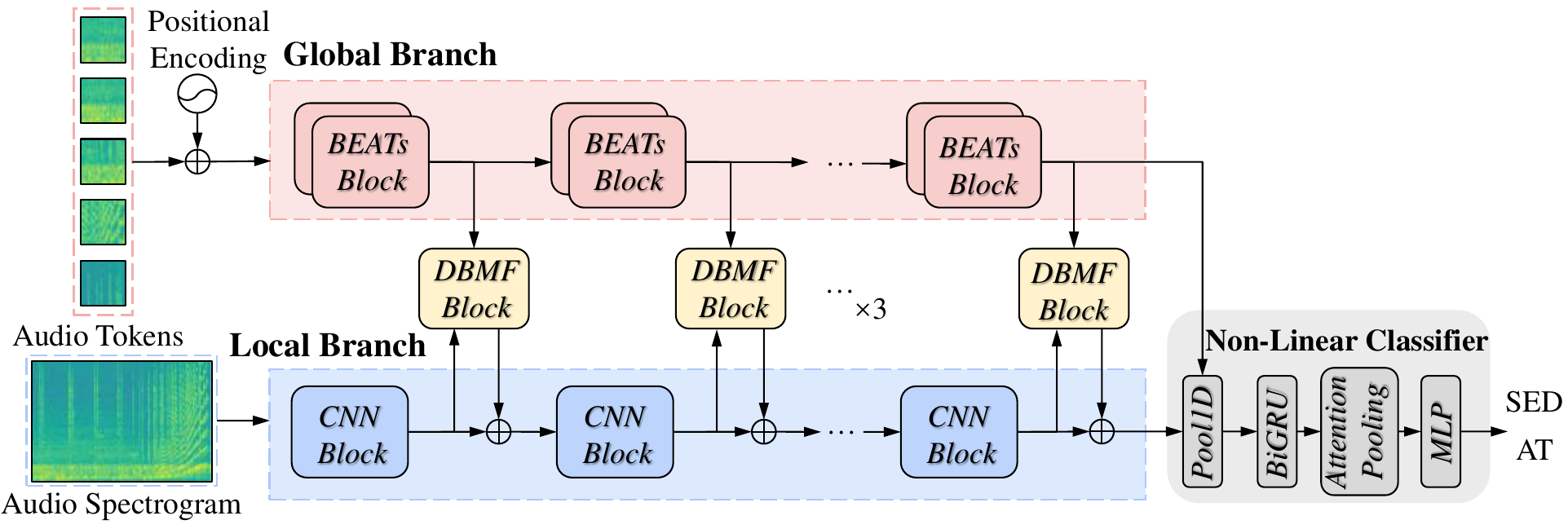}%
        \label{fig:pipeline}
    }
    \caption{Architecture Overview of the MTDA-HSED: We provide a comprehensive visualization of the MTDA-HSED, divided into two main components. Fig 1(a) illustrates the M3A module, showcasing the integration of Long-Term Audio Adapter and Short-Term Audio Adapter within the BEATs block’s FFN. Fig 1(b) depicts the dual-branch pipeline comprising the BEATs, CRNN, and DBMF module.}
    \label{fig:method}
\end{figure*}

We categorize the challenges of utilizing heterogeneous dataset in SED into two main aspects: \textbf{scenario discrepancy} and \textbf{granularity discrepancy}. 1) \textbf{scenario discrepancy}: Different datasets are collected in different ways, leading to discrepancy in the scenario of subsets within heterogeneous dataset. These excessive scenarios present significant challenges for models in robustly learning a wide range of scenarios. 2) \textbf{granularity discrepancy}: Under the supervision of the soft and hard labels, the model learns the two patterns of features, respectively, with an inherent granularity discrepancy between them. This discrepancy poses a challenge that SED systems need to prudentially accommodate both granularities' information when encoding features. Otherwise, it might lead to the loss of partial key information about sound events and the confusion of both granularities' information. In other fields, some methods have demonstrated success, such as ACT-NET~\cite{yoo2022enrichedcnntransformerfeatureaggregation}, which have addressed the challenge of accommodating both granularities in the field of image super-resolution through layer-by-layer fusion.

In this paper, we propose an effective specialized architecture named \textbf{M}utual-Assistance \textbf{T}uning and \textbf{D}ual-Branch \textbf{A}ggregating for \textbf{H}eterogeneous \textbf{SED} task (MTDA-HSED). Specifically, to address the above challenges, we propose two novel modules. Firstly, we introduce a novel additive tuning method: \textbf{M}utual-\textbf{A}ssistance \textbf{A}udio \textbf{A}dapter (M3A). The M3A is inserted into the Feed Forward Network (FFN) of the BEATs block, which effectively fine-tunes the model by modifying the pre-trained structure, enhancing its ability to capture comprehensive feature information, and coping with the multi-scenario nature of SED with heterogeneous dataset. Secondly, to interact with the coarse-grained information and fine-grained information, we propose a novel aggregation approach, the \textbf{D}ual-\textbf{B}ranch \textbf{M}id-\textbf{F}usion module (DBMF). This module, inserted between the BEATs and CNN branches, facilitates the deep integration of global information from the BEATs branch and local information from the CNN branch. It can avoid the loss of partial key information about sound events and the confusion of both granularities' information, thereby enhancing the performance of the detection.

\section{Methodology}

\subsection{Mutual-Assistance Audio Adapter}\label{AA}

Compared to additive methods, existing fine-tuning approaches in SED preserve the original structure of pre-trained models, thereby limiting their ability to tackle the challenge of models in robustly learning a wide range of scenarios. Therefore, we propose a novel additive tuning method named M3A to promote the robustness of the model for multi-scenario applications, as shown in Fig.~\ref{fig:M3A}. 

The M3A has a symmetrical structure, which contains a Long-Term Audio Adapter and a Short-Term Audio Adapter. Specifically, the Long-Term Audio Adapter employs an inverted bottleneck structure to project the input feature into a high-dimensional space and processes the projection with ReLU function to extract high-richness information before restoring it to its original dimension. This high-richness information is a mixture that contains not only time-frequency pattern of sound event but also adjacent information of different events. The Short-Term Audio Adapter employs a bottleneck structure to extract low-richness information using an identical approach, where the input feature is projected into a low-dimensional space and processed using the Softmax activation function. This low-richness information contains only the time-frequency patterns of the sound event itself, with less attention paid to the adjacent information. Thus the Short-Term Audio Adapter tackles the time-frequency patterns better than the Long-Term Audio Adapter when faced with the disjointed short-term sound event. Specifically, the M3A structure can be formulated as follows.

\begin{equation}
    x^{\prime}_\ell = \mathrm{LN}(attn + x_{\ell - 1})
    \label{eq:M3A_1}
\end{equation}
\begin{equation}
    x^{\prime \prime}_{\ell,1} = \mathrm{ReLU}(x^{\prime}_{\ell} \cdot \bm{W}_{up}) \cdot \bm{W}_{down}
    \label{eq:M3A_2}
\end{equation}
\begin{equation}
    x''_{\ell, 2} = \mathrm{Softmax}(x'_{\ell} \cdot \bm{W}_{down}) \cdot \bm{W}_{up}
    \label{eq:M3A_3}
\end{equation}
\begin{equation}
    x''_{\ell} = s_1 \cdot x''_{\ell, 1} + s_2 \cdot x''_{\ell, 2} + \mathrm{FFN}(x'_{\ell})
    \label{eq:M3A_4}
\end{equation}
\begin{equation}
    x_\ell = \mathrm{LN}(x''_{\ell} + x'_{\ell})
    \label{eq:M3A_5}
\end{equation}
where $x_{\ell - 1} \in \mathbb{R}^{T \times D}$ is the input of the BEATs block from $\ell - 1$ layer, $x'_{\ell}$ is the input of the M3A-FFN in the $\ell$ layer, $x''_{\ell, 1}$, $x''_{\ell, 2}$ and  $x''_{\ell}$ are the output of Long-Term Audio Adapter, Short-Term Audio Adapter and the M3A-FFN respectively, $attn \in \mathbb{R}^{T \times D}$ is the output of the Multi-Head Self-Attention, $\bm{W}_{down}$ and $\bm{W}_{up}$ are the parameters of down-projection layer and up-projection layer respectively, $s_1$ and $s_2$ are the scaling parameters for mutual-assistance, $x_{\ell}$ is the output from the $\ell$ layer.

\subsection{Dual-Branch Mid-Fusion Module}

The official baseline lacks a deep fusion of global information and local information, leading to ineffectively addressing the challenge in prudentially accommodating both granularities' information when encoding features. Therefore, it is crucial to interact deeply with BEATs and CNN branches.

\begin{figure}
    \centering
    \includegraphics[scale=0.38]{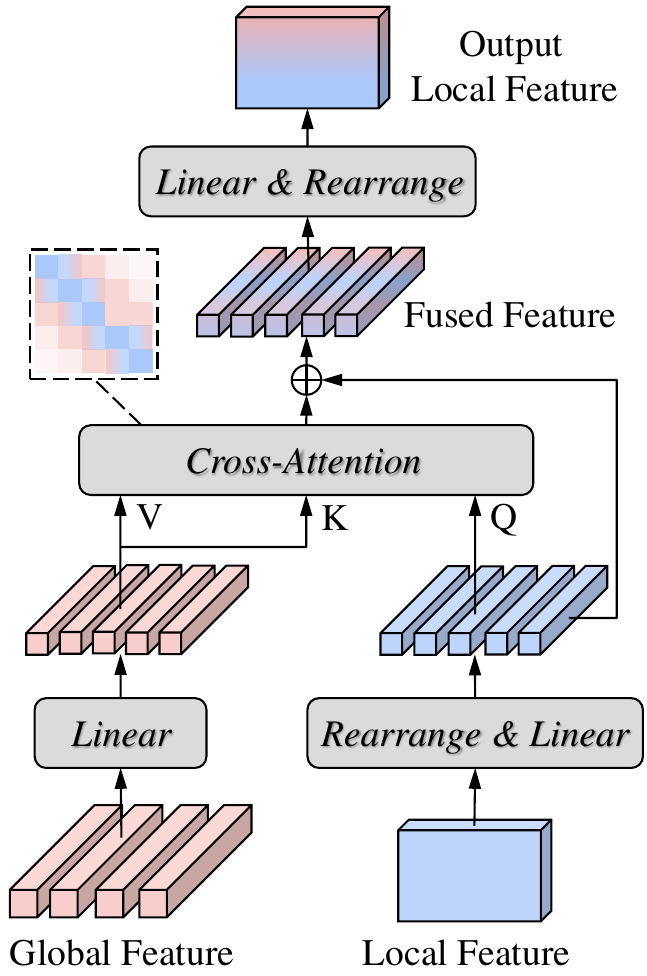}
    \caption{Details of the DBMF Module}
    \label{fig:enter-label}
\end{figure}

As shown in Fig.~\ref{fig:pipeline}. In our pipeline, global feature from the pre-trained BEATs branch and local feature from the CNN branch are first mapped into a common embedding space via a linear layer. And then they are fed into the DBMF module simultaneously. The output of the DBMF module is combined with the local feature in a residual way, and finally, the combination result is sent to the next CNN block. As for the design of the DBMF module, we employ a Multi-Head Cross-Attention mechanism, where the global feature $g$ is input as key and value, and the local feature $l$ is input as a query, as shown in Fig.~\ref{fig:enter-label}. In this way, global information from the BEATs branch can be incorporated into local feature from the CNN branch gradually over layers. Consequently, a deep aggregation feature of the two branches can be obtained, which contains comprehensive information across diverse data granularities. This process can be expressed using the following formulas.

\begin{equation}
    g^{\prime} = \mathrm{Linear}(g)
    \label{eq:DBMF_1}
\end{equation}
\begin{equation}
    l^{\prime} = \mathrm{Linear}(\mathrm{Rearrange}(l))
    \label{eq:DBMF_1}
\end{equation}
\begin{equation}
    f = \mathrm{Softmax}\left(\frac{(l^{\prime} \cdot \bm{W}_{q})(g^{\prime} \cdot \bm{W}_{k})^{T}}{\sqrt{d}}\right)g^{\prime} \cdot \bm{W}_{v}
\end{equation}
\begin{equation}
    f^{\prime} = \mathrm{FFN}(\mathrm{LN}(f)) + f
\end{equation}
\begin{equation}
    f^{\prime \prime} = \mathrm{Linear}(\mathrm{Rearrange}(f^{\prime}))
\end{equation}
where $g \in \mathbb{R}^{T_1 \times D_g}$ and $l \in \mathbb{R}^{c \times t \times f}$ are the features of global and local branches, $g' \in \mathbb{R}^{T_1 \times D}$ and $l' \in \mathbb{R}^{T_2 \times D}$ are the features after linearization. $\mathrm{Rearrange}$ represent the process of sequencing the local feature, $f \in \mathbb{R}^{T_2 \times D}$ represents the output after feature aggregation, $\bm{W}_{q}, \bm{W}_{k}, \bm{W}_{v} \in \mathbb{R}^{D \times d}$ are learnable parameters and $d$ is the dimension of learned query, key and value vectors. Note that global-local feature aggregation occurs every two BEATs blocks and every CNN block.

\section{Experiment}

\subsection{Implementation Details}
DESED \cite{Turpault2019_DCASE, Serizel2020_ICASSP} dataset and MAESTRO Real \cite{Martinmorato2023} dataset are chosen to conduct experiments. 
The baseline model is the BEATs-CRNN architecture, and means teacher (MT)~\cite{Tarvainen17} is applied for consistency training with unlabeled data for semi-supervised learning. 
Data augmentations such as mixup~\cite{zhang18}, time masking~\cite{park19} are used.
The data augmentation and metrics hyper-parameters are identical to those of the baseline. 

Methods are evaluated with polyphonic sound event detection scores (PSDS1)~\cite{Bilen20}, PSDS1(sed score)~\cite{ebbers2022thresholdindependentevaluationsound} and mean pAUC (mpAUC). The mpAUC is computed only on MAESTRO dataset (and on MAESTRO classes), while the PSDS1 and PSDS1(sed score) are computed only on DESED dataset (and on DESED classes). That means PSDS1 and PSDS1(sed score) reflect on the evaluations with hard labels for sound events, while the mpAUC reflects on the evaluations with soft labels for sound events.

\begin{table}[htbp]
\setlength{\tabcolsep}{3pt}
\centering
\caption{Comparison of related innovation methods. The best results are in \textbf{bold}, and the second best are in \underline{underlined}. PSDS1(s) means PSDS1(sed score).}
\label{table 1}
\begin{tabular}{ccccc}
\toprule
\textbf{Model} & \textbf{Type}& \textbf{PSDS1↑} & \textbf{PSDS1(s)↑} & \textbf{mpAUC↑} \\
\midrule
Baseline   & -& 0.494 & 0.499 & 0.709 \\

\midrule

ATST-SED \cite{shao2023finetunepretrainedatstmodel} &selective &0.297& 0.301& 0.554 \\

MONA \cite{yin2023adapterneedtuningvisual} &additive& 0.497& 0.507&0.709  \\
ADAPTER \cite{chen2022adaptformeradaptingvisiontransformers} &additive& 0.494& 0.503&0.704  \\

ACT-NET \cite{yoo2022enrichedcnntransformerfeatureaggregation} & fusion& 0.308& 0.316&0.696  \\

M3A(ours)&additive& \underline{0.503}& \underline{0.511}& \underline{0.753} \\

DBMF(ours)&fusion& 0.494& 0.501& 0.748 \\

MTDA-HSED(ours)&additive and fusion& \textbf{0.503}& \textbf{0.514}& \textbf{0.757} \\
\bottomrule
\end{tabular}
\end{table}

\subsection{Experiment Results}
\subsubsection{Performance Comparison}
We compare the baseline method with different fine-tuning and interaction methods. The fine-tuning methods include ATST-SED, MONA, ADAPTER, and our M3A; the interaction methods include ACT-NET and our DBMF.

The results are shown in Table~\ref{table 1}. It is clear that our MTDA-HSED method outperforms various other methods on three metrics. In addition, it can be seen that both the proposed M3A and DBMF modules outperform other methods in their respective fields. Firstly, in the field of fine-tuning, the additive methods include MONA~\cite{yin2023adapterneedtuningvisual}, ADAPTER~\cite{chen2022adaptformeradaptingvisiontransformers}, and M3A are significantly better than selective methods, including the ATST-SED ~\cite{shao2023finetunepretrainedatstmodel}. It is possible to prove that the additive methods, by altering the structure of BEATs, is able to better tackle SED task with heterogeneous dataset. In the additive methods, the proposed M3A module shows its superiority, which outperforms MONA~\cite{yin2023adapterneedtuningvisual}and ADAPTER~\cite{chen2022adaptformeradaptingvisiontransformers} in PSDS1, PSDS1(sed score) and mpAUC. The results indicate that the M3A is effective in the face of SED with heterogeneous dataset. Secondly, in the area of interaction, our DBMF method outperforms the baseline method and outperforms ACT-NET~\cite{yoo2022enrichedcnntransformerfeatureaggregation} by at least 5\% on all metrics, which proves that the proposed DBMF module interacts deeply with the global and local features, avoidance of the confusion of both granularities’ information and effective preservation of partial key information about sound events, thereby improving the detection performance.  

\vspace{-10pt}
\begin{table}[htbp]
\centering
\caption{Comparison of different audio adapter number, N.}
\label{table 2}
\begin{tabular}{cccc}
\toprule
\textbf{N} &\textbf{PSDS1↑} & \textbf{PSDS1(sed score)↑} & \textbf{mpAUC↑} \\
\midrule
1& 0.494& 0.503&\underline{0.740}  \\
2& \textbf{0.503}& \textbf{0.511}& \textbf{0.753} \\
3& \underline{0.499}& \underline{0.508}& 0.706 \\
\bottomrule
\end{tabular}
\end{table}

\begin{table}[htbp]
\centering
\caption{Comparison of different projection dimension(The scaling factor based on the original data), D.}
\label{table 3}
\begin{tabular}{cccc}
\toprule
\textbf{D} &\textbf{PSDS1↑} & \textbf{PSDS1(sed score)↑} & \textbf{mpAUC↑} \\
\midrule
2,1/2 & 0.487& 0.499&\textbf{0.757} \\
4,1/4& \textbf{0.503}& \textbf{0.511}& 0.753 \\
\bottomrule
\end{tabular}
\end{table}

\subsubsection{Ablation Study}

\paragraph{\textbf{Impact of audio adapter number and projection dimension in M3A}}

In the ablation experiments related to the M3A module, we perform the experiments with different number of audio adapter and projection dimension, respectively.

In the ablation experiments with different number of audio adapter, as shown in Table~\ref{table 2}, the performance of setting the number to two is better than other audio adapter numbers. This demonstrates that in the face of the heterogeneous dataset containing two subsets, we need to design symmetrical audio adapters to make BEATs learn a wide range of information. 

The results of ablation experiments with different projection dimensions for the Long-Term Audio Adapter and Short-Term Audio Adapter are shown in Table~\ref{table 3}. The best results are achieved when scaling the original feature dimension to $4x$ and $1/4x$, respectively. This illustrates that a significantly different projection dimension is helpful in extracting different richness information in heterogeneous dataset, which is critical for the model to learn the scenario discrepancy of subsets within heterogeneous dataset.

\paragraph{\textbf{Impact of aggregate strategy}}

In the section on ablation experiments with the DBMF module, we perform experiments with different aggregate strategies. Specifically, we experiment with different streams between the dual branches, including single-stream BEATs to CNN (B$\rightarrow$C), single-stream CNN to BEATs (C$\rightarrow$B), and bidirectional-stream (C$\leftrightarrow$B).

As mentioned above, the PSDS1 and the PSDS1(sed score) focus on fine-grained information, while the mpAUC focus on coarse-grained information. The results can be found in Table~\ref{table 4}. In single-stream, the B$\rightarrow$C are slightly worse than C$\rightarrow$B in the face of capturing the fine-grained information, while the B$\rightarrow$C outperforms C$\rightarrow$B in the face of capturing the coarse-grained information. This could be attributed to the B$\rightarrow$C process, which focuses on global information and marginalizes the local information in the aggregated feature, thus enhancing performance in evaluations with hard labels. Compared to the single-stream experiments, the results from the bidirectional-stream experiment are significantly worse. It is possible that after the bidirectional interaction, the aggregated feature may lose the original and distinctive information from the dual branches. 

\begin{table}[htbp]
\centering
\caption{Comparison of different streams.}
\label{table 4}
\begin{tabular}{cccc}
\toprule
\textbf{Stream} & \textbf{PSDS1↑} & \textbf{PSDS1(sed score)↑} & \textbf{mpAUC↑} \\
\midrule
B $\rightarrow$ C& \underline{0.494}& \underline{0.501}&\textbf{0.748}  \\
C $\rightarrow$ B& \textbf{0.499}& \textbf{0.508}&0.544  \\
C $\leftrightarrow$ B& 0.435& 0.442& \underline{0.556} \\

\bottomrule
\end{tabular}
\end{table}

\subsubsection{Qualitative Study}

\begin{figure}
    \centering
    \includegraphics[scale=0.38]{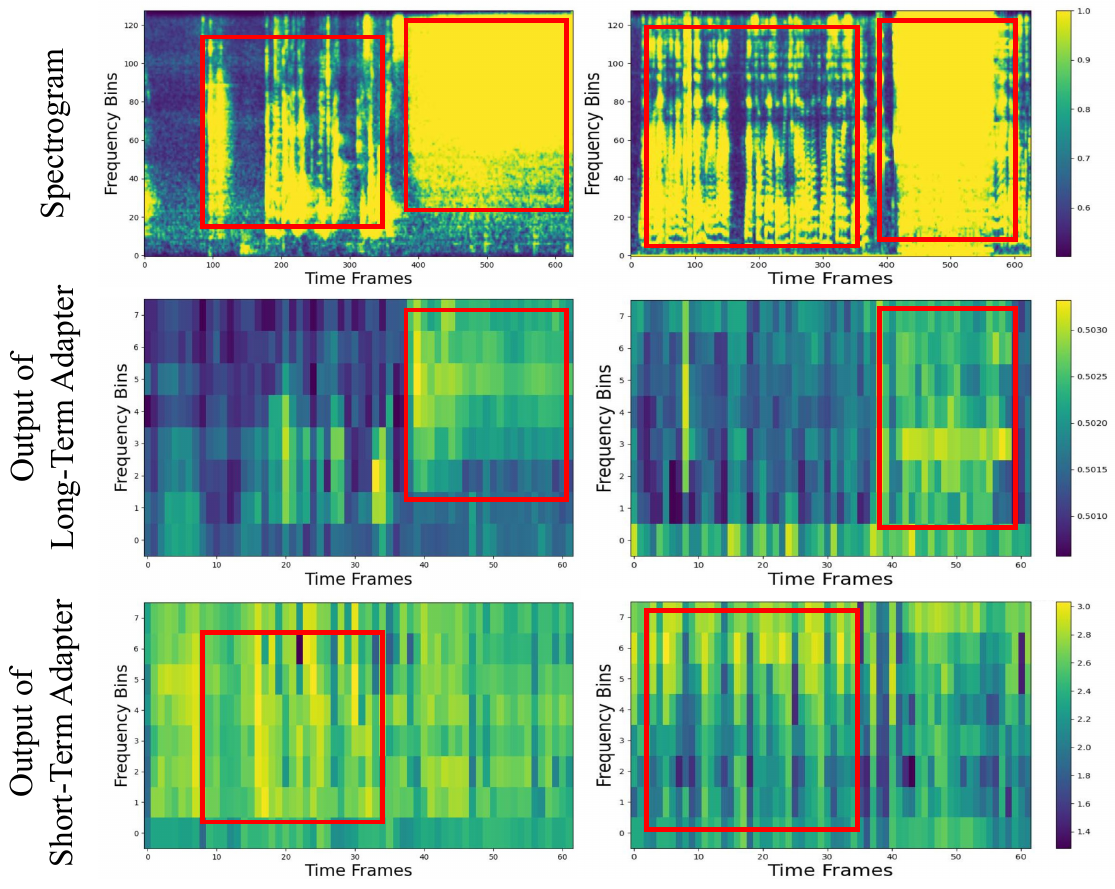}
    \caption{Visualization of the M3A modules: The first line is the spectrogram of the input. The second line is the output of the Long-Term Audio Adapter. The third line is the output of the Short-Term Audio Adapter.}
    \label{fig:adapter}
\end{figure}

The visualization results are shown in Fig.~\ref{fig:adapter}. By comparing the feature maps of the Long-Term and Short-Term Audio Adapter with the spectrograms of the input data, we can see that the time-frequency patterns modeled by the Short-Term Audio Adapter and the Long-Term Audio Adapter are diverse. It may be due to the fact that the Long-Term Audio Adapter expands the feature information by up-projection and introduces the adjacency information of adjacent events. This adjacency information enriches the information of the long-term event, as shown in the second line of Fig.~\ref{fig:adapter}, and fills the interval between the disjoint short-term event, so that the Long-Term Audio Adapter focuses more on the long-term information. Inversely, the Short-Term Audio Adapter ignores adjacent information by down-projection, which makes it more specialized in modeling the time-frequency patterns of the short-term sound events, as shown in the third line of Fig.~\ref{fig:adapter}.

\vspace{-5pt}
\section{Conclusion}
In this paper, we introduce the MTDA-HSED architecture, which is an effective architecture with dual branches for SED task with heterogeneous dataset. In MTDA-HSED, we propose two novel modules. Firstly, the M3A module is an additive method for fine-tuning the BEATs by inserting a symmetrical structure with different projection dimensions into the FFN of BEATs block, which aims to cope with the multi-scenario nature of SED with heterogeneous dataset. Secondly, the DBMF module is a mid-fusion method to interact with global and local feature obtained by the BEATs and CNN branches with the cross-attention mechanism, which aims to combine multi-granularity information and avoid the loss of partial key information about sound events. The experimental results demonstrate the effectiveness of the proposed method.

\section*{Acknowledgment}

This work was supported partly by the National Natural Science Foundation of China (Grant No. 62173045, 62273054), partly by the Fundamental Research Funds for the Central Universities (Grant No. 2020XD-A04-3), and the Natural Science Foundation of Hainan Province (Grant No. 622RC675).

\footnotesize
\bibliographystyle{IEEEtran}
\bibliography{reference}

\end{document}